\documentclass[iop]{emulateapj}

\usepackage{graphicx}
\usepackage{epsfig}
\usepackage{txfonts}
\usepackage{amssymb}
\usepackage{epsf}
\usepackage{natbib}

\newcommand{\vc}[1]{\mathbf{#1}}
\newcommand{\ipsf}{ISP}

\submitted{}
\accepted{May 11, 2010}

\shorttitle{Ionospheric Power-Spectrum Tomography in Radio Interferometry}

\shortauthors{Koopmans}

\begin{document}

\title{Ionospheric Power-Spectrum Tomography in Radio Interferometry}

\author{L.V.E.\ Koopmans}

\affil{Kapteyn Astronomical Institute, University of Groningen, P.O.Box 800, 9700 AV Groningen, The Netherlands ({\tt koopmans@astro.rug.nl})}

\begin{abstract}
A  tomographic method is described to quantify the three-dimensional power-spectrum of the 
ionospheric electron-density fluctuations based on radio-interferometric observations by a 
two-dimensional planar array. The method is valid to first-order Born approximation and might 
be applicable to correct observed visibilities for phase variations due to the imprint of 
the {\sl full} three-dimensional ionosphere. It is shown that not the ionospheric electron density distribution is 
the primary structure to model in interferometry, but its autocorrelation function or equivalent its power-spectrum. 
An exact mathematical expression is derived that provides the three dimensional power-spectrum of the ionospheric electron-density 
fluctuations directly from a rescaled scattered intensity field and an incident intensity field convolved with 
a complex unit phasor that depends on the $w$-term and is defined on the full sky pupil plane. In the limit of a  small field of view, 
the method reduces to the single phase screen approximation. Tomographic self-calibration can
become important in high-dynamic range observations at low radio frequencies with wide-field 
antenna interferometers, because a three-dimensional ionosphere causes a spatially
varying convolution of the sky, whereas a single phase screen results in a spatially invariant convolution.
A thick ionosphere can therefore not be approximated by a single phase screen without introducing errors
in the calibration process. By applying a Radon projection and the Fourier projection-slice theorem, it is shown 
that the phase-screen approach in three dimensions is identical to the tomographic method. 
Finally we suggest that residual speckle can cause a diffuse
intensity halo around sources, due to uncorrectable ionospheric phase fluctuations 
in the short integrations, which could pose a fundamental limit on the dynamic range in long-integration images. 
\end{abstract}


\section{Introduction}

Over the last years, considerable effort has gone into the construction of large wide-field low-frequency 
radio interferometers. One of many challenges that these new instruments must overcome
is that of phase and amplitude fluctuations due to the ionosphere \citep{1951RSPSA.209...81H, 1952RSPSA.214..494H}. This is particularly
important at meter-wavelegths where these instruments operate \citep[see e.g.][]{2009AJ....138..439C}
and the refractive index of the ionized medium can be
particularly large\footnote{The ratio of the plasma over the observing frequency at 150 MHz is $\sim$0.03; hence intensity 
scattering is at the level of $\sim$0.001 and scales with $\lambda^{2}$.}. Moreover, low-frequency radio  telescopes often 
have wide fields of view (i.e.\ up to tens of degrees), 
making them not only susceptible to the combined electric field from
large parts of the sky, but also to its modifications by the full three-dimensional electron-density structure of the ionosphere. 

For high-frequency interferometers -- having dish-antennae that probe a small area of the sky (a few degrees or less) -- the 
three-dimensional ionosphere can be collapsed along the line-of-sight and be well approximated by a two-dimensional 
thin phase-screen, without loosing substantial information \citep[]{1967ApJ...147..433S}. This approximation, however, does not hold if the field of 
view becomes large and widely separated antennae observe different sources, under large angles, through similar 
parts of the ionosphere \citep[e.g.][]{2005ASPC..345..399L}. In the case of wide-field low-frequency interferometers, a full three dimensional model of the ionosphere
is required for self-consistent ionospheric calibration.
Whereas considerable work has recently gone in to developing (multi-layer) two-dimensional wide-field ionospheric models 
\citep[e.g.][]{2009A&A...501.1185I, 2009arXiv0911.3942M}, no physically intuitive and self-consistent description seems readily available for full-scale three-dimensional wide-field modelling of the ionosphere. 

In this paper a step towards such a complete physical description is presented based on the tomographic theory of electric field scattering by weakly
inhomogeneous media \citep[]{1969OptCo...1..153W}, as applied to the ionosphere. The main result is that the ionosphere, over a wide field of view, acts  
as a scatterer with a {\sl spatially varying} point spread function. The instantaneous
two-dimensional point spread function at a position $\vc{s}$ (a directional unit vector) around a point source at $\vc{s}_{0}$ is identical to the instantaneous three-dimensional power-spectrum\footnote{Throughout the paper, if not mentioned otherwise, the power-spectrum is 
assumed to be the exact squared amplitude of the ionospheric electron density waves as function of their three dimensional wave 
vectors {\sl at a given moment in time}, 
and not the expectation value of this squared amplitude over an ensemble average \citep[][]{1981A&A...102..212B}. In general the former is a Gaussian random realization of the latter.} of the ionospheric electron-density 
sampled from points $(\vc{s}-\vc{s}_{0})$ on the surface of an ``Ewald sphere of reflection'' \citep[][]{1969AcCrA..25..103E}, i.e.\ a spherical surface with unit radius centered on the
direction of the source. 
Having many point or very compact sources over a larger field of view allows one to sample the three-dimensional
Fourier structure of the ionosphere at many distinct points. These points can be used to constrain a three dimensional 
model of the power-spectrum of the ionosphere. This model can subsequently be used to ``deconvolve'' the image, also at points
where there are no strong sources, and remove the 
dominant phase errors due the three dimensional ionosphere. We
term this ``tomographic self-calibration'', because it involves a three dimensional ionosphere and not a single phase screen. This calibration can not be done in the classical way through self-calibration \citep[see e.g.][]{1984ARA&A..22...97P}
because the scattering point spread function is not spatially invariant as is implicit in that method. 
It either requires the solution of a matrix 
(measurement) equation \citep[e.g.][]{1996A&AS..117..137H, 2009arXiv0911.3942M, 2010arXiv1001.5268L}, which is computationally
more demanding than traditional self-calibration or, as we will show, it can be written as a specific three dimensional Fourier 
transform. 

The main goal of this paper, however, is not to introduce a specific solution scheme or algorithm to these equations, but to provide 
more useful guidance to future methods of self-consistent {\sl physics-motivated} three 
dimensional modeling of the ionosphere and its use in calibration. 
The second goal is to describe several effects of a three dimensional 
ionosphere on interferometric measurements at low frequencies that go beyond thin 
phase-screen models, which have thus far been very successful, but are demonstratively incorrect for an extended
three dimensional ionosphere.

The outline of the paper is as follows:\ In Section 2, the theory of scattering of a plane-wave electric field is restated 
in terms more familiar to radio interferometry and extended to the case of multiple point sources.
In Section 3, the cross-correlation of the scattered electric field is determined (i.e.\ the visibility function) and several effects of 
the ionosphere on imaging are described (e.g.\ speckle). In Section 4, the first order effect of the thickness of the ionosphere is 
analyzed. In Section 5, the description is extended 
from multiple points sources to a continuous intensity field and a mathematical expression is derived that allows one to build
a three dimensional model of the power spectrum of the ionosphere from information obtained
only in the interferometer plane (i.e.\ a holographic principle). In Section 6, it is shown how the tomographic method,
connects to an extension of the phase-screen approach to three dimensions using a Radon transformation and applying 
the Fourier projection-slice theorem.  In Section 7, we summarize our results and give conclusions. 

\section{Weak Scattering by the Ionosphere}

In this section, the basic theory of electric-field tomography of a weakly scattering media is restated as 
first introduced by \citep[]{1969OptCo...1..153W}. The notation is adapted to that typically used
in radio astronomy and radio interferometry \citep[see e.g.][]{2001isra.book.....T}.
Under the assumption that the refractive index of the ionosphere varies slowly over a single photon wavelength,
and the ionosphere is ``frozen'' over the time scale the radiation passes through it, one can write the electric field equation in a dielectric medium as 
\begin{equation}
	\nabla^{2}\vc{E}_{\nu}(\vc{r}) + [k^{2} n_{\nu}^{2}(\vc{r})]\, \vc{E}_{\nu}(\vc{r})=0,
\end{equation}
where $n_{\nu}(\vc{r})$ is the refractive index of the medium at a position ${\vc r}$ and 
at frequency $\nu$, and $k=2\pi/\lambda$ is the usual wave number for a wavelength $\lambda = c/\nu$
\citep[][]{1999prop.book.....B}. 
In a plasma with electron density $n_{e}(\vc{r})$ the refractive index is given by 
$n^{2}={1-[n_{e} e^{2}/(\nu^{2} m_{e} \epsilon_{0})]}\equiv {1-(\nu_{\rm p}/\nu)^{2}}$, 
where $ \nu_{\rm p}$ is the plasma frequency, being typically $\sim$5\,MHz for the ionosphere.
We drop the explicit frequency dependence of the 
electric field, but it should implicitly be assumed.

Because the three components of the electric field are independent, each
component satisfies the same solution and we can simply use the scalar $E$ to describe the electric field. 
If the refractive index of the ionosphere is near unity, which is often the case in radio astronomy (i.e.\
$\nu \gg  \nu_{\rm p}$), the equation for the electric field can be conveniently rewritten as
\begin{equation}
	(\nabla^{2}  + k^{2})\, E(\vc{r}) = - 4 \pi\, \Phi(\vc{r}) \, E(\vc{r}),
\end{equation}
where $\Phi(\vc{r})\equiv k^{2}[n^{2} - 1]/4 \pi$ is called the {\sl scattering potential}.
In case of $n\approx1$, the scattering potential strength is close to zero and scattering is weak\footnote{We note here that even though scattering can be ``weak'' in terms of a small deviation of the refractive index from unity everywhere, the integral over the line of sight can still lead to substantial phase changes in the electric field, leading to both diffractive and refractive effects and strong scintillation \citep[e.g.][]{1992RSPTA.341..151N}. In the current discussion ``weak scattering'' means both a refractive index very close to 
unity {\sl and} phase fluctuations much less than unity. The latter is the nominal mode of the ionosphere and data obtained with interferometers during strong scintillation (e.g.\ during solar bursts or sun rise/set) occur only rarely and are often discarded in analyzes.}. When 
$\Phi(\vc{r}) = 0$ everywhere, the equation reduces to the Helmholtz equation for a plane wave.
The solution of this equation can be obtained \citep[][]{1999prop.book.....B} through Green's functions and leads to an implicit integral equation
\begin{equation}
	E(\vc{r}) = E^{(i)}(\vc{r}) + \int_{V} \Phi(\vc{r}') E(\vc{r}') \frac{e^{i k |\vc{r} - \vc{r}'|}}{|\vc{r} - \vc{r}'|} d^{3}\vc{r}',
\end{equation}
where $E(\vc{r}) = E^{(i)}(\vc{r}) + E^{(s)}(\vc{r})$. The first term is the incident (plane) wave and the second term
the scattered wave. The latter is equal to the integral term above and carried out over the 
entire volume $V$ in which $n\neq 1$. The last factor in the integrand indicates that $E^{(s)}(\vc{r})$ 
is a spherically outgoing wave, assuming the extent of the scattering potential is small compared to the
distance between observer and scattering medium. 

\subsection{Weak Scattering of Multiple Point Sources}

We now extend the single plane-wave description of \citet[]{1969OptCo...1..153W} to an incident electric field that 
results from the sum of $N$ point sources that satisfy the solution of the Helmholtz equation in free space. 
Later in the paper we further extend this to a continuous intensity field. We also express all physical
distances in units of $\lambda$, i.e. $\vc{u} \equiv \vc{r}/\lambda = (u,v,w)$ in the remainder of the paper. 
Its Fourier equivalent is $\vc{s}=(s_{u},s_{v},s_{w})$.
The incident electric field in this case becomes
\begin{equation}
	E^{(i)}(\vc{u}) = \sum_{n} \sqrt{S_{n}}\, e^{2 \pi i \vc{s_{0,n}} \cdot \vc{u}},
\end{equation}
where $S_{n}$ is the flux-density of point source $n=1\dots N$ and $\vc{s}_{0,n}$ are unit vectors that point in the 
directions of the point sources. 
These point sources are (for now) assumed to dominate the electric field and can be compared to the {\sl 
phase calibrators} in radio interferometry.
In the weak scattering approximation, the scattered wave has a relatively low amplitude compared to the incident wave. 
We can then replace $E(\vc{u})$ with $E^{(i)}(\vc{u})$
to  first-order (Born) approximation, finding
\begin{equation}
	 E_{1}^{(s)}(\vc{u}) = \sum_{n} \sqrt{S_{n}} \int_{V} \Phi(\vc{u}')  e^{2 \pi i \vc{s_{0,n}} \cdot \vc{u}' } \frac{e^{2 \pi i |\vc{u} - \vc{u}'|}}{|\vc{u} - \vc{u}'|} d^{3}\vc{u}',
\end{equation}
where the subscript indicates that the scattered field is that to first order Born approximation
and $\Phi(\vc{u}) \equiv [n^{2}-1]$. 
We note here that the second order can 
be obtained by substituting $E^{(i)}(\vc{u}) + E_{1}^{(s)}(\vc{u})$ in to the original integral equation to obtain a solution
to second order, etc. This iterative scheme only works for weak scattering, where the values of 
$\Phi(\vc{u})$ do not exceed unity.

If we use the fact that the spherical outgoing wave can be written as \citep[][]{1919AnP...365..481W}
\begin{equation}
\frac{e^{2\pi i |\vc{u} - \vc{u}'|}}{|\vc{u} - \vc{u}'|}  = i {\int_{-\infty}^{\infty}\int_{-\infty}^{\infty}}  \frac{1}{s_{w}} e^{2 \pi i (s_{u} (u-u') + s_{v} (v-v') + s_{w} (w-w')} ds_{u} ds_{v},
\end{equation}
with $s_{w}^{2}={1-s^{2}_{u}+s^{2}_{v}}$ for $s^{2}_{u}+s^{2}_{v}\le1$ or
$s_{w}^{2}=- ({s^{2}_{u}+s^{2}_{v}-1})$ otherwise. Complex values of $s_{w}$ lead to  exponentially decaying 
(evanescent) electric fields that are typically not measurable far from the scatterer. 
The other (homogeneous) waves are those measured by a distant observer.
If one further uses the Fourier transform of the scattering potential,
$\tilde{\Phi}(\vc{s}) = \iiint \Phi(\vc{u}) e^{-2 \pi i \vc{s} \cdot \vc{u}} d^{3}\vc{u},$
one can write the scattered electric field as
\begin{equation}
	E_{1}^{(s)}(\vc{u}) = i \sum_{n} \sqrt{S_{n}} \iint  \frac{1}{s_{w}} \tilde{\Phi}(\vc{s} - \vc{s}_{0,n})\
		e^{2 \pi i \vc{s}\cdot \vc{u}} ds_{u} ds_{v},
\end{equation}
where we assume a geometry where $w=0$ is the ground-plane below the ionosphere where $n=1$, 
and that $w>0$ is in the direction of the zenith or the phase reference center (see below). The interferometer is placed in a plane defined at a 
constant $w_{\rm ant} = z_{\rm ant}/\lambda$. Typically one can assume $w_{\rm ant}=0$. 

Thence, one finds a relation between the Fourier transform of the observed electric field in the plane of the interferometer at $w_{\rm ant}$ and the Fourier transform of the scattering potential
\begin{equation}\label{eqn:scattered_field}
	\tilde{E}^{(s)}(s_{u}, s_{v}) = \frac{i}{s_{w}}  e^{+ 2\pi i s_{w} w_{\rm ant}} \sum_{n} \sqrt{S_{n}} \tilde{\Phi}(\vc{s} - \vc{s}_{0,i}), 
\end{equation}
with 
$	\tilde{E}^{(s)}(s_{u}, s_{v}) = \iint E_{1}^{(s)}(u,v,w_{\rm ant}) e^{ + 2 \pi i (s_{u} u + s_{v} v)} du dv$.
This can be regarded as the 
Fourier transform of a two-dimensional slice through a three-dimensional scattered electric field. In this 
paper we do not treat the case of an interferometer with varying $w_{\rm ant}$. A planar array is an reasonable assumption for relatively compact (i.e.\ km-scale) interferometers, but breaks down
on large scales where the curvature of the Earth can not be neglected \citep[see e.g.][]{2009MNRAS.395.1558C}. For a planar array, however, the $w$-term due to the array can be neglected for small integration times (i.e.\ instantaneous
sampling of the electric field in a plane), in contrast to visibilities from very different time frames where the array has rotated over a substantial
angle compared to the phase center (only a linear east-west array does not suffer from the $w$-term). 

The physical interpretation of Eqn.(\ref{eqn:scattered_field}) is the following: Every point of the two-dimensional Fourier transform of the scattered electric field in the plane 
of an interferometer probes a single three-dimensional mode of the scattering potential (i.e.\ the scattering 
medium) for a single point source. In the presence of $N$ point sources, all in different directions, every point of the two-diminsional 
Fourier transform of the scattered electric field in the plane of an interferometer probes the sum of  $N$ independent three-dimensional 
modes of the scattering potential. In Section 5 we show how to unravel this information.

\section{The Effects of Weak Ionospheric Scattering on Interferometric Images}

In radio interferometry one does not analyze the electric field itself. In that case, Eqn.(\ref{eqn:scattered_field}) would directly 
yield the three-dimensional structure of 
the ionosphere (per integration time) because the phase information of the Fourier transform of the 
electron density of the ionosphere is fully retained
in the phase information of the scattered electric field. In reality, only the 
cross-correlations of the electric field, measured at different antennae pairs, are stored (i.e.\ the complex visibilities)
and the phase information of the ionospheric density fluctuations is lost. 
In the following, we assume that the total electric field from the entire sky (i.e.\ the antenna sensitivity 
is directionally independent) is measured over the infinite interferometer plane with  $w=w_{\rm ant}$. 

Visibilities are sampled from the cross-correlation of the electric field $E(\vc{u}) = E^{(i)}(\vc{u}) + E^{(s)}(\vc{u})$
with its complex conjugate, i.e.\ $V(\vc{b})\equiv \langle E(\vc{u}) E^{*}(\vc{u} + \vc{b}) \rangle_{\rm t}$ with 
$\vc{b}$ being the baseline between two points (antennae) in plane of the interferometer. The averaging is assumed
to be over time. The Fourier transform 
of the visibilities forms the incident intensity from the sky, as follows from the van Cittert-Zernike theorem
\citep[e.g.][]{2009MNRAS.395.1558C}. The same intensity is also the product of the Fourier transform of the electric field 
with its complex conjugate. 
A bit of algebra shows that 
the cross-correlation between the incident and scattered fields depends on
the imaginary part of the zero-mode, $\tilde{\Phi}(0)$, of the ionosphere, and
consequently is equal to zero. 
The multiplication of the Fourier transform of the scattered electric field with its complex conjugate therefore
provides the complete scattered intensity 
\begin{equation}
	\delta I^{(s)}(s_{u}, s_{v})= \langle \tilde{E}^{(s)*}(s_{u}, s_{v}) \tilde{E}^{(s)}(s_{u}, s_{v}) \rangle_{\rm t},
\end{equation} 
where  the dependence on $w_{\rm ant}$ disappears. Using Eqn.(\ref{eqn:scattered_field}), we find the following result
\begin{equation}
	\delta I^{(s)}_{\rm coh}(s_{u}, s_{v})= \frac{1}{s_{w}^{2}} \sum_{n} \sum_{m} \sqrt{S_{n} S_{m}} 
		\tilde{\Phi}^{*}(\vc{s} - \vc{s}_{0,n})  \tilde{\Phi}(\vc{s} - \vc{s}_{0,m}).
\end{equation}
This equation is exact for phase-coherent point sources to first order Born approximation. However,
the sky is an incoherent emitter \citep[see][for an exposure on the coherence properties of electric fields]{1965RvMP...37..231M}. Hence, the cross-terms with $n\neq m$ depend on the electric field coming from 
incoherent point sources and vanish, such that we are left with 
\begin{equation}\label{eqn:scattering_psf}
	\delta I^{(s)}(s_{u}, s_{v})= \frac{1}{s_{w}^{2}} \sum_{n}  S_{n} \,
		|\tilde{\Phi}(\vc{s} - \vc{s}_{0,n}) |^{2},
\end{equation}
where we dropped the subscript. 
This equation forms the basis for further discussions in the paper. The above equation
is only correct for an interferometer and an electric field measured in a plane. In three dimensions, one would
no longer be able to use simple Fourier transforms (see below), because $s_{w}$ depends explicitly on $s_{u}$ and $s_{v}$.

To understand the physical interpretation of the above equation, one might suppose a point source in the zenith (or 
equivalently in the phase center) emitting a plane wave in the absence of the ionosphere. Because the phase of the electric field is the same
at each antenna (by construction), its Fourier transform yields a complex delta function in the zenith with a time-varying
phase. Multiplied
with its complex conjugate, this recovers the point source intensity. If a two-dimensional thin phase-screen 
is placed in between the source and the array, exhibiting a single wave-mode 
in electron density perpendicular to the zenith or phase reference center direction,  
then part of the electric field amplitude will be modulated such that its phases show to first order the imprint of 
this ionospheric wave-mode \citep[see][for a wonderful description]{1956RPPh...19..188R}. 
The modulated phase (i.e.\ a single wave over the array) can be interpreted as being identical in the weak scattering limit to the modulated 
phase of a point source 
offset from the zenith in the direction of the ionospheric wave-vector by a distance set by the phase-frequency over the array. 

Hence, {\sl a single ionospheric wave-mode scatters a fraction of a point source intensity into an additional 
point source (a speckle), offset by the projected phase frequency onto the array and with an intensity proportional 
to the amplitude of the wave-mode squared}. The sum of all speckles create a halo of scattered emission around
the point source, when not corrected for through phase calibration.

\subsection{Speckle and Speckle Noise}

Because the phase fluctuations of the electric field are measured in the plane of the array and are typically
drawn from a Gaussian random field realization from some (ensemble average) power spectrum, 
taking the Fourier tranform of this field and multiplying it with its complex conjugate yields that each point source 
exhibits a diffuse ``halo'' of scattered intensity \citep[e.g.][]{1967ApJ...147..433S, 1972ApJ...174..181C}. 
Its Fourier transform is related to the usual phase structure function ($D_{\phi}$; see below). This pattern is referred to 
as ``speckle'' in optical (laser) interferometry. We define the instantaneous ionospheric 
scattering PSF (\ipsf\ hereafter) as the sum of a delta function plus its scattered speckle pattern,  re-normalized to
a flux-density of unity such that flux is conserved in a convolution process. 
The \ipsf\ is related to the Fourier transform of optical transfer function \citep[][]{1985stop.book.....G}, determined by 
phase fluctuations induced by the ionosphere. 
Eqn.(\ref{eqn:scattering_psf}) shows that the point-source intensity is convolved with an \ipsf\ that 
reflects a curved surface in the Fourier transform of the instantaneous ionospheric electron-density fluctuations.  
Disregarding a geometrically determined distortion that depends only on the angle of the 
observed point on the sky away from the zenith, a three-dimensional ionosphere causes a spatially 
varying convolution. In contrast, a thin two-dimensional ionospheric screen causes a spatially invariant
convolution. This sets our analysis apart from most studies up to the present that have focussed
on scattering by ionized media \citep[see e.g.][for an excellent review]{1992RSPTA.341..151N} where wide-field
effects and extended (thick) screens can be neglected in nearly all circumstances.

For ionospheric electron-density fluctuations set by Kolmogorov turbulence, one expects that $|\tilde{\Phi}(\Delta \vc{s}) |^{2} \propto |\Delta \vc{s}|^{-\beta}$ with $\beta= 
11/3$ over a scale of meters to tens of kilometers \citep[e.g.][]{2001isra.book.....T}. This implies that in the weak scattering regime, point sources on average exhibit a speckle pattern which rapidly decreases 
in intensity with distance from the source \citep[][]{1985stop.book.....G}. The speckle pattern, at a given moment in time, is a Gaussian random realization 
from a power-spectrum with expectation value $|\tilde{\Phi}(\Delta \vc{s}) |^{2}$. 
Scattered
flux is spread over scales from arcseconds to degrees, corresponding to tens of km to tens of meter-scale ionospheric electron density 
fluctuations, respectively. It is interesting to note that ionospheric modes of a gives physical scale give rise to image distortions
that are visible only on baselines equal or larger to that same physical scale. The properties of large isoplanatic patches over a given scales are therefore easier to determine if one can observe the sky
with an array that exceeds that scale \citep[see also e.g.][]{1992A&A...257..401J}. Long baselines thus significantly help in calibrating shorter baselines, even if the
science of interest is obtained on the short baselines\footnote{To correct for all modes in the field of view ($\theta_{\rm FOV}$), the longest baseline should 
be $b \ga 25\,{\rm km} \times [H_{\rm ion}/(300 {\rm ~km})] \cdot [\theta_{\rm FOV}/({\rm 5~deg})]$, where $H_{\rm ion}$ is the typical 
height of the ionoshere.}. 

The presence of a speckle pattern can have consequences. Once the \ipsf\ reaches an intensity level comparable to 
the average intensity of the surrounding sources, the confusion noise, or the noise itself, one can no longer distinguish it 
from these contaminants and calibration is typically limited to a maximum $k$-mode. This could place a limit
on the calibratibility of the instrument, especially at very low frequencies near the ionospheric cutoff,  because the signal 
to noise ratio of 
an image snapshot, taken over the maximum time-scale of substantial ionospheric
changes, might not be large enough to correct for low signal to noise speckles further away from the sources. This leaves
residual speckles which are a function of wavelength (scaling with $\lambda^{2}$ in 
strength and with $\lambda$ in scale). 

A detailed calculation of the residual speckle intensity and its associated noise goes goes beyond the scope of the current 
paper.
We note that these speckles do {\sl not} average 
away over time but ultimately form a halo (i.e.\ ``seeing'') around each source. Whereas a bright source might be very
useful in determining the ionospheric scattering, it might also leave a residual speckle pattern in the image due
to uncalibrated  phase fluctuations from ionospheric scale below the smallest baseline. It could therefore be 
advantageous to place bright sources
near the half-power of the field of view such that they still probe the same ionosphere to first order, but that their residual
speckle patterns do not contaminate the field of interest.

Speckle noise (i.e. the expected standard deviation from the expectation value of the speckle intensity at a given point), however, does average away. To see this, we note that
a convolution of the intensity implies a multiplication of the visibilities with the Fourier transform of the \ipsf. In the case
of weak scattering and assuming a thin phase screen and small field of view, this multiplication function to first order 
is $1-D_{\phi}(\theta)$ where the latter term is the phase structure function. \citet[][]{1972ApJ...174..181C}
showed that the variance in the scattered part of the intensity is set by
the amplitude of $D_{\phi}(\theta)$ is to first order equal to the its expectation value. Hence speckle noise
around point sources average away as one over the square root of the number of independent realizations of the ionosphere.
Obviously calibration can remove much of the speckle pattern, but not that caused by the structure below the 
smallest and above the longest baselines or below the noise level, respectively.
This leftover halo of speckle can be compared to that in optical (after AO correction) in the search of faint companions around nearby stars \citep[e.g.][]{1999PASP..111..587R}. Speckle and speckle noise ultimately set the detection threshold and a similar effect can 
play a role in very high dynamic range imaging at low radio frequencies. We finally note that a convolution implies a suppression of visibilities on long baselines. In particular, if no calibration is applied for ionospheric effects, phases on long baselines becomes less
correlated and average out stronger \citep[][]{1954RSPSA.225..515B} and a ``seeing'' halo develops around point sources.

\section{Variation in the Ionospheric Scattering PSF over the Field of View}

In the case of a single source in the phase-center and narrow field of view over which $s_{z}$ changes little from zero,  
the \ipsf\ around a 
point source (assuming a perfectly sampled uv-plane or electric field) represents 
the instantaneous power-spectrum of the electron density fluctuations on a two-dimensional 
surface in three-dimensional Fourier space. In the case of a narrow FOV and a compact 
speckle pattern with $s_{w} \rightarrow 0$, this pattern is nearly equal to 
the instantaneous two-dimensional power spectrum of the medium integrated along the line-of-sight, which 
has thus far motivated the use of a phase-screen or thin ionospheric model \citep[e.g.][]{2009A&A...501.1185I}.

To assess how the \ipsf\ is modified when displaced
from the phase reference center \citep[see][usually the center of the antenna beam]{2001isra.book.....T}, 
we assume that the  phase reference
center  is in the zenith, such that the 
$w$--axis points exactly upward and the $uv$--plane is perpendicular to it. The planar array
lies at constant $w$ as assumed thus far. It can be
shown, however, that for any planar array, the coordinate system can be rotated to any phase
reference center. By correcting the phases of each of the antennas such that the fringe-rate is
zero for a chosen phase reference center, one effectively re-orients the planar array such that it 
acts as being perpendicular to the line connecting the array and the phase-reference center. Hence, the analysis
performed in the coordinate system with the phase reference center in the zenith is in fact 
valid for any chosen phase-reference center for planar arrays by defining the $w$-axis in that direction.

We now assume two point sources, one in the phase reference center  at $\vc{s}_{\rm zenith}=(0,0,1)$ and one 
slightly offset from the zenith at a unit vector $\vc{s}_{\rm src}$, such that $|\vc{s}_{\rm src}-\vc{s}_{\rm zenith}|\ll1$.
One can then Taylor expand the \ipsf\ to first order. To do this properly, however, one needs to compare two
points that are offset from the two sources (on in the zenith and one offset from it) by an identical two-dimensional vector $\delta\vc{s}_{\rm 2D}=(\delta s_{u}, \delta s_{v})$. In that case,
one compares the ratio of their {\sl aligned} speckle patterns in the $(s_{u},s_{v})$ plane. 
Furthermore, because the first two coordinates of 
$\vc{s}-\vc{s}_{\rm src} = (\delta\vc{s}_{\rm 2D}, \delta s_{w}-\delta s^{\rm src}_{z}=(1-|\delta\vc{s}_{\rm 2D}|^{2})^{1/2} - (1-|\delta \vc{s}^{\rm src}_{\rm 2D}|^{2})^{1/2})$ are the same by construction for both sources, any difference occurs because of the dependence
of the \ipsf\ on $\delta s^{\rm src}_{w}$ and hence one only needs to Taylor expand Eqn.(\ref{eqn:scattering_psf}) for a single 
point source with 
respect to changes in $\delta\vc{s}^{\rm src}_{\rm 2D}$, which is the offset of the source from the phase-center. 

After a bit of algebra one finds that the fractional intensity difference between the 
offset \ipsf\ divided by the one in the phase center is
\begin{eqnarray}\label{eqn:fraction}
	f^{(s)}_{I}(\delta\vc{s}_{\rm 2D}) &\approx & \frac{2\,\delta \vc{s}^{\rm src}_{\rm 2D}\cdot \delta\vc{s}_{\rm2D}}{1-|\delta\vc{s}_{\rm2D}|^{2}} \left[
		1-\frac{\sqrt{1-|\delta\vc{s}_{\rm2D}|^{2}}}{2 |\tilde{\Phi}(\vc{t}) |^{2}} \frac{\partial {|\tilde{\Phi}(\vc{t}) |^{2}}}{\partial t_{z}}\right] 
\end{eqnarray}
with the derivative of the power-spectrum being evaluated at the point $\vc{t}=(\delta\vc{s}_{\rm 2D}, \sqrt{1-|\delta\vc{s}_{\rm 2D}|^{2}}-1)$. The length of $\delta\vc{s}_{\rm2D}$ does not necessarily have to be much smaller than unity
because so far we only expanded in $\delta \vc{s}^{\rm src}_{\rm 2D}$. 
If we further assume that also $|\delta\vc{s}_{\rm2D}| \ll 1$ then the equation further simplifies to
\begin{eqnarray}
	f^{(s)}_{I}(\delta\vc{s}_{\rm 2D}) &\approx & 2\,\delta \vc{s}^{\rm src}_{\rm 2D}\cdot \delta\vc{s}_{\rm2D} \left[
		1-\frac{\partial {\ln |\tilde{\Phi}(\vc{t}) |^{2}}}{ 2\, \partial t_{z}}\right].
\end{eqnarray}
The \ipsf\ is therefore no longer 
spatially invariant but starts to depend on the 
power-spectrum of the ionosphere in the $w$-direction and on the offset of the source from the phase reference center. 
Whereas to first order the \ipsf\  remains identical to that of a thin phase-screen,  the large-scale structure (i.e.\ with $t_{w}\approx 0$) of the ionosphere
in the $w$-direction becomes important because ${\partial {|\tilde{\Phi}(\vc{t}) |^{2}}}/{\partial t_{z}} \neq {\rm constant}$ for a thick ionosphere. Because this additional term depends
strongest on the large scale $w$-modes, we expect this term to be rather insensitive to relative small changes in $\vc{s}^{\rm src}_{\rm 2D}$ and show slower diurnal and seasonal variations. 

If, instead of working with the image $\delta I^{(s)}(s_{u}, s_{v})$, we work with $\delta J^{(s)}(s_{u}, s_{v}) = s_{w}^{2}\ \delta I^{(s)}(s_{u}, s_{v})$ (see also
Section 5.2), the fractional difference is 
\begin{eqnarray}\label{eqn:fraction_rescaled}
	f^{(s)}_{J}(\delta\vc{s}_{\rm 2D}) &\approx & -({\delta \vc{s}^{\rm src}_{\rm 2D}\cdot \delta\vc{s}_{\rm2D}}) \left[
	\frac{\sqrt{1-|\delta\vc{s}_{\rm2D}|^{2}}}{|\tilde{\Phi}(\vc{t}) |^{2}} \frac{\partial {|\tilde{\Phi}(\vc{t}) |^{2}}}{\partial t_{z}}\right] \nonumber \\
		&\approx & -({\delta \vc{s}^{\rm src}_{\rm 2D}\cdot \delta\vc{s}_{\rm2D}}) \left[
	 \frac{\partial {\ln |\tilde{\Phi}(\vc{t}) |^{2}}}{\partial t_{z}} \right]
\end{eqnarray}
A physical interpretation of this is the following: when observing away from the phase center, one
becomes sensitive to large scale modes in the $w$-direction. This causes a gradient in the electron density 
over a scale that corresponds to depth of the ionosphere along the line of sight to the source projected onto the 
array. For small angles away from the phase center  this projected baseline is small and interference between
the opposite parts of the ionosphere is not symmetric. This results in large scale interference patterns on the
plane of the array. These patterns correlated strongest on long baselines and thus show up as small scale structure
in the \ipsf, i.e.\ near the point sources. We note that in the case of a thin
ionosphere, which has ${\partial \ln |\tilde{\Phi}(\vc{t}) |^{2}}/{\partial t_{w}}\equiv 0$, the \ipsf\ is spatially invariant
to all orders. 

The interpretation of Eqn.(\ref {eqn:fraction_rescaled}) is that offsets,  $\vc{s}_{\rm 2D}$, perpendicular to $\vc{s}^{\rm src}_{\rm 2D}$
lead to a zero fractional difference, whereas parallel offsets lead to maximal differences (either positive of negative). This implies that to lowest order the \ipsf\
is invariant in the tangential direction when moved from the phase center on radial spokes
and is stretched or squeezed in the radial direction further away from the phase-center, depending on the gradient of the ionospheric 
power-spectrum in the $w$-direction. Eqn.(\ref{eqn:fraction_rescaled}) can thus be used to determine the \ipsf\ over a wider field of view, taking the
first-order effect of the thickness of the ionosphere into account. By having several calibrators spread over the field of view,
one can determine the 
nearly constant value of ${\partial \ln |\tilde{\Phi}(\vc{t}) |^{2}}/{\partial t_{w}}$. Once known, the \ipsf\ can be 
determined to lowest order for the rest of the field of view by simple rescaling.

To estimate the effect of the thickness of the ionosphere let us assume it has a thickness $d_{\lambda}$ in units
of wavelength and a uniform electron density in the $w$-direction. The normalized power-spectrum
in $s_{w}$ direction has the functional form $|\tilde{\Phi}(t_{w}) |^{2} = (d_{\lambda}/\pi)\times {\rm sinc}^{2}(d_{\lambda} t_{w})$.
Inserting this in Eqn.(\ref{eqn:fraction_rescaled}) and Taylor expanding to lowest order yields, along radial spokes, 
$f^{(s)}_{J} \approx (\pi/3)\ d_{\lambda}\ |\delta\vc{s}_{\rm 2D}|^{3}\ |\delta\vc{s}^{\rm src}_{\rm 2D}|$.
We note that even though the fractional changes in the \ipsf\ are smaller near the sources\footnote{
Numerically we find  $f^{(s)}_{J} \sim 10^{-2}\, (d_{\lambda}/10^{5})\ |\delta\vc{s}_{\rm 2D}/{10'}|^{3}\ |\delta\vc{s}^{\rm src}_{\rm 2D}/{10^{\circ}}|$. As example, for an ionosphere with 200\,km thickness and observed at 2 meter wavelength, 
the \ipsf\ about 10$'$ away from sources near the edge of a 10 degree field of view changes by $\sim$1\% of the \ipsf\ maximum. This might
seem small, but could still be considerably larger than the noise or confusion level for bright sources in the field. Hence, in high 
dynamic range and wide field of view imaging experiments the third dimension of the ionosphere becomes very important.}, the absolute changes are larger because
of the steep increase of the power spectrum, i.e.
$ |\tilde{\Phi}(\delta \vc{s}_{\rm 2D}) |^{2} \times f^{(s)}_{J} \propto |\delta\vc{s}_{\rm 2D}|^{-3/2}\ |\delta\vc{s}^{\rm src}_{\rm 2D}|$ 
for $\beta=11/3$ i.e.\ a Kolmogorov spectrum. Note that this equation breaks down near the 
outer scale (i.e.\ whose effect shows up closest to the images) of the ionosphere otherwise it would not be bounded.

\section{Three dimensional Ionospheric Power-Spectrum Tomography}

After having derived an equation in the previous section that describes the scattering of point sources due to the three-dimensional
ionosphere, here we introduce the equivalent expression for a continuous intensity
field and then discuss the consequences of these results for ionospheric calibration.

\subsection{Scattering of a Continuous Intensity Field}

One can simply extend the point-source equation for scattering (Eqn.(\ref{eqn:scattering_psf})) 
to a convolution-type operation on 
a continuous intensity field by relating flux-density with intensity times area. In that case one readily sees that
\begin{equation}\label{eqn:cont_field}
	\delta I^{(s)}(s_{u},s_{v})= \frac{1}{s_{w}^{2}} \iint  I^{(i)} (s_{0,u},s_{0,v})\,
		|\tilde{\Phi}(\vc{s} - \vc{s}_{0}) |^{2} ds_{0,u} ds_{0,v}.
\end{equation}
We note that this is a two-dimensional convolution with a three dimensional kernel. This makes
it more difficult to deconvolve using simple two-dimensional Fourier techniques. For a small field of
view, one can simply set $s_{w}-s_{0,w}=0$ and perform a two dimensional convolution through 
fast Fourier transform methods. For wider fields of view this can not be done. However, one can rewrite the 
equation as a three dimensional convolution as follows
\begin{eqnarray}\label{eqn:cont_field2}
	\delta I^{(s)}(s_{x},s_{y}) &=& \frac{1}{s_{w}^{2}} \iiint  I^{(i)}_{\rm 3D} (\vc{s}_{0})\,
		\delta_{\rm k}(\delta s_{w}) |\tilde{\Phi}(\vc{s} - \vc{s}_{0}) |^{2} d\vc{s}_{0},
\end{eqnarray}
where $\delta s_{w} = s_{0,w}-(1-s_{0,u}^{2}-s_{0,v}^{2})^{1/2}$ and $\delta_{\rm k}$ is a Kronecker delta function.
Note that in this equation $\vc{s}_{0}$ should no longer be treated as a unit vector but still that $(1-s_{0,u}^{2}-s_{0,v}^{2})\ge 0$.
The intensity $I^{(i)}_{\rm 3D} (\vc{s}_{0})$ is
a cylinder of radius unity that has the same value as $I^{(i)}(s_{0,u},s_{0,v})$ for each value of $s_{w}$. 
Writing the equation like this, makes it a three dimensional convolution which can be performed using Fourier transform techniques, but at the cost of more memory and computational effort. 

\subsection{Determination of the Three-Dimensional Ionospheric Power Spectrum}

Here we derive an expression for the full ionospheric power spectrum in terms of the 
scattered intensity field and the incident field. To do this, first we define the rescaled version of the scattered intensity as
\begin{equation}\label{eqn:rs_scattering}
	\delta J^{(s)}(s_{u},s_{v}) \equiv s_{w}^{2} \cdot \delta I^{(s)}(s_{u},s_{v}),
\end{equation}
and  $\delta \tilde{J}^{(s)}(u,v) = \iint \delta J^{(s)}(s_{u},s_{v}) e^{+ 2\pi i  (s_{u} u+ s_{v} v)} ds_{u} ds_{v}$ as its Fourier transform. We note that the scattered intensity is zero if $s_{u}^{2}+s_{v}^{2}>1$.
If we now define \citep[see e.g.][for a similar approach for non-coplanar arrays\footnote{Only after having introduced this
transformation, the author became aware of a similar transformation in these publications. Whereas the later assume a three-dimensional (i.e.\ non-planar) array, the focus in the current paper is deriving the structure of the three-dimensional (non-planar) ionosphere with a
planar array. Combining both is left for a future publication.}]{1989ASPC....6..117S, 1992A&A...261..353C}
\begin{equation}
	J^{(i)}_{\rm 3D}(\vc{s}_{0}) \equiv I^{(i)}_{\rm 3D} (\vc{s}_{0})\cdot
		\delta(s_{0,w}-\sqrt{1-s_{0,u}^{2}-s_{0,v}}),
\end{equation}
we obtain
\begin{equation}
	\delta J^{(s)}(s_{u},s_{v})=  \iiint  J^{(i)}_{\rm 3D}(\vc{s}_{0}) \, |\tilde{\Phi}(\vc{s} - \vc{s}_{0}) |^{2} d\vc{s}_{0}.
\end{equation}
We note that this equation can be integrated over infinity, as long as the intensities are zero (as they are) when
$|\vc{s}_{0}|>1$. Using the 
relation between convolution and Fourier transforms, we can now write this
as
\begin{equation}\label{eqn:FT_sf}
	\delta \tilde{J}^{(s)}(u,v)=  \tilde{J}^{(i)}_{\rm 3D}(\vc{u}) \cdot {\cal F}(|\tilde{\Phi}(\vc{s}) |^{2})(\vc{u})
\end{equation}
with $\tilde{J}^{(i)}_{\rm 3D}(\vc{u}) =  \iiint {J}^{(i)}_{\rm 3D}(\vc{s}_{0}) e^{- 2 \pi i \vc{s}_{0} \cdot \vc{u}} d\vc{s}_{0}$ and ${\cal F}(|\tilde{\Phi}(\vc{s}) |^{2})$ being the autocorrelation function of the 
ionospheric scattering function. This remarkable equation shows that the two dimensional 
field $\delta \tilde{J}^{(s)}(u,v)$ contains information about the full three dimensional structure of the ionosphere 
if a reference 
field $\tilde{J}^{(i)}_{\rm 3D}(\vc{u})$ is available; this is closely related to the technique of ``holography''.
We can now go one step further and show after a little algebra that
\begin{equation}
	\tilde{J}^{(i)}_{\rm 3D}(\vc{u}) = \tilde{I}^{(i)}(u,v)\ \otimes\ {\cal H}(u,v;w).
\end{equation}
being a two-dimensional convolution, with $w$ as control parameter. The function
\begin{equation}\label{eqn:Hankel}
{\cal H}(u,v;w) = 2 \pi \int_{0}^{1} e^{-2 \pi i w \sqrt{1-s^{2}}} J_{0}(s\cdot u_{\rm 2D})\ s \ ds, 
\end{equation}
with $u_{\rm 2D} = \sqrt{u^{2}+v^{2}}$ is a Hankel transform of $e^{-2 \pi i w \sqrt{1-s_{u}^{2}-s_{v}^{2}}}$.
Putting this all together, we find the final result
\begin{equation}\label{eqn:powerspectrum}
	|\tilde{\Phi}(\vc{s}) |^{2} = {\cal F}^{-1}\left[\frac{\delta \tilde{J}^{(s)}(u,v)}{\tilde{I}^{(i)}(u,v)\ \otimes\ {\cal H}(u,v;w)}\right].
\end{equation}
Hence, the three dimensional power spectrum of the ionosphere or its autocorrelation function can be reconstructed from the 
ratio between the Fourier transform of the rescaled scattered intensity field and the Fourier transform of the 
incident radiation field convolved with a Hankel function. 

{\sl So what does Eqn.(\ref{eqn:powerspectrum}) mean?} Looking carefully at the equation, we see that ${\cal H}(u,v;w)$ is the Fourier transform 
of a unit phasor with a phase that is determined by the distance (in wavelength) from a half-sphere of radius 
$w$ to the $uv$ plane along a line perpendicular to the latter. 
In other words, ${\cal H}(u,v;w)$ act as the Fourier transform of a complex optical transfer function in the pupil plane
\citep[see e.g.][]{1985stop.book.....G},  which in this case is the full sky and not (as usual) the interferometer plane itself. Equivalently, 
${\cal H}(u,v;w)$ acts as a complex point spread function in $uv$ space (i.e.\ convolving $\tilde{I}^{(i)}(u,v)$). 
Turning this around, a convolution in $uv$
space is a multiplication of the sky intensity with the reciprocal of the convolution kernel. 
Hence, ${\cal H}(u,v;w)$ causes a complex beam of unit amplitude on the sky equal to the phasor in Eqn(\ref{eqn:Hankel}). 
This exactly extracts the information from $\delta \tilde{J}^{(s)}(u,v)$ on
a particular $w$-slice cut through the three dimensional autocorrelation function of the ionosphere. Hence
by simply multiplying the sky model ${I}^{(i)}(s_{u},s_{v})$ with the complex phasor in Eqn.(\ref{eqn:Hankel}) one 
obtains a complex sky-intensity cube. Fourier transforming this back slice by slice provides the 
denominator in Eqn.(\ref{eqn:powerspectrum}).  

To illustrate this further, imagine that the sky contains only one point source of unit flux-density at $(s_{0,u}, s_{0,v})$. In that case, 
\begin{equation}\label{eqn:ps2}
	{\cal F}(|\tilde{\Phi}(\vc{s}) |^{2}) = \left[\frac{\delta \tilde{J}^{(s)}(u,v)}{e^{-2 \pi i (s_{0,u} u + s_{0,v} v + w \sqrt{1-s_{u,0}^{2}+s_{v,0}^{2}})}}\right].
\end{equation}
Substituting Eqn.(\ref{eqn:cont_field2}) back into this equation shows that the left and right-hand sight of the equation
are identical as required. By bringing the denominator to the left side of the equation, we find for a point source
\begin{equation}\label{eqn:ps3}
	\left|\tilde{\Phi}(\vc{s}-\vc{s}_{0}) \right|^{2} = {\cal F}^{-1} \left[{\delta \tilde{J}^{(s)}(u,v)}\right],
\end{equation}
This is the inversion of Eqn.(\ref{eqn:scattering_psf}) for a single point source. Hence every point source in the sky probes the ionospheric power spectrum on an Ewald sphere of reflection. The full sky probes the power spectrum of the ionosphere, as encoded in Eqn.(\ref{eqn:powerspectrum}), on the surface of many offset Ewald spheres of reflection, each of unit radius.
The function ${\cal H}$ endows  each point in the sky with a complex
phase. In correspondence with the Fourier shift theorem (i.e.\ a phase shift corresponds to a spatial shift) this point is then moved in to the third dimension 
of power-spectrum which corresponds to the $w$-direction of the ionosphere\footnote{
The phase in the phasor changes by $2\pi$ when $w s \delta s \approx 1$. If the maximum of $s$ 
is set by the field of view and $\delta s \sim \lambda/u_{\rm max}$ by the resolution of the array, the heigh $w$ to which the ionosphere can 
be probed is $w_{\rm max} \approx 1/(s \delta s)$ above which phases wrap around
multiple times inside a resolution
element in the image (i.e.\ information is being lost). Taking a typical field of view of $\sim$5 degrees and resolution of
3 arcmin (for an array of 2-3 km diameter) leads to a maximum high of $w_{\rm max} \lambda \approx 25,000$\,m for $\lambda = 2$\,m. Structure at ten times that height, where the ionosphere typically is densest, is encoded in the image 
on scales ten times below the resolution limit. Hence, baselines well beyond 20-30 km are required to fully extract 
the three dimensional structure of the ionosphere and its effect on the scattering in the image. One also finds that the scale over which the \ipsf\ will
change is $s \sim u_{\rm max}/w_{\rm max}$. This is typically tens of degrees for the ionosphere and arc-minutes in the optical.}. 

\subsection{The effects of the beam, $uv$ sampling}

Realistic interferometers neither observe the full sky, nor sample the electric 
field in the full plane of the array. If we define the complex electric field beam pattern as ${B}(s_{u},s_{v})$ and
the sampling function of the electric field on the plane of the interferometer (i.e.\ usually delta functions at the 
positions of the antennae) as ${P}(u,v)$ and its Fourier transform as $\tilde{P}(s_{u},s_{v})$, then 
Eqn.(\ref{eqn:powerspectrum}) for incoherent emitters can be rewritten as:
\begin{equation}
	\delta \tilde{J}^{(s)} = \left[{\cal F}(|\tilde{\Phi} |^{2})\times {\cal F}(|\tilde{P}|^{2}) \times (\tilde{I}^{(i)}\ \otimes\ {\cal H})\right]\ \otimes\ {\cal F}\left(\frac{|B|^{2}}{s^{2}_{w}}\right), 
\end{equation}
where we suppressed the explicitly dependence of the functions on the $(s_{u},s_{v},s_{w})$ or $(u,v,w)$. The equation holds for each given $w$. 
The usual two dimensional array point spread function (i.e.\ dirty beam) is denoted by $|\tilde{P}|^{2}$ and its Fourier
transform is the visibility sampling function in the $uv$ plane. Similarly, $|B|^{2}$ is the usual two dimensional antenna beam pattern. This equation shows clearly that the visibilities of the incident intensity
field are both multiplied with the Fourier transforms of power-spectrum of the ionosphere and that of the $uv$ sampling 
function, respectively. The resulting field is then convolved with the Fourier transform of the rescaled beam pattern,
which to first order (if the beam is much smaller than the full sky) is the aperture gain pattern of the antennae. To derive $|\tilde{\Phi} |^{2})$
from this equation thus requires deconvolution of the observed visibilities of the scattered intensity field. This could be 
difficult in principle were it not for the fact that the beam size in general is much larger than the extent of the \ipsf\ and 
the dirty beam. In that case the scale over which ${\cal F}(|\tilde{\Phi} |^{2})\times {\cal F}(|\tilde{P}|^{2})$ varies is very large
compared to the convolution kernel and the latter can be neglected for modes much smaller than the field of view. 
Hence in practice it is expected that the deconvolution is 
not really required to obtain an accurate evaluation of the power-spectrum (i.e.\ $|\tilde{\Phi} |^{2}$).

\section{Connection to the Phase Screen}

Whereas in this paper we started our discussion from the Born approximation and derived the scattered field, 
this is only a valid approach in the weak scattering scattering or weak scintillating regime. Despite this approximation,
it is remarkably accurate up to intensity fluctuations very close to unity.
However, there are other ways to solve for the scattered electric field \citep[e.g.][for a detailed review]{1992RPPh...55...39K}
in the weak scattering regime that are closely connected to the formalism as introduced above and to the phase-screen
approach. Instead of looking at every point of the medium as a source of a single scattering (as is done in the Born 
approximation), one can also assume that the medium does not modify the amplitude of the incident wave to first order and 
that light rays travel on a straight line through the medium. In that case, since the medium does not absorb or amplify 
the wave, only a phase shift occurs between different points in the medium when a plane wave enters the ionosphere.

We can again describe this in terms of the refractive index or electron density as follows. A phase shift of $\delta \psi \approx 
k \int \delta n(\vc{r})\, ds$ is introduced between a wave traveling in a medium with refractive index $1+\delta n(\vc{r})$ and unity,
respectively, where the integral is carried out along a straight line through the medium \citep[see e.g.][]{1971ApOpt..10.1652L}. A plane wave of a source with unit amplitude entering medium exits as a wave with a ``wrinkled'' phase-front due to varying refractive indices along different paths \citep[][]{1972ApJ...174..181C}. The auto-correlation of this ``phase-screen'' is the function with which the visibilities of that source are multiplied in the $uv$-plane and 
its Fourier transform is the \ipsf\ as we discussed it before. This allows us to directly connect the previous analysis with 
that of the phase-screen through the so-called Radon transform, which is related to the Fourier projection-slice theorem.

The Fourier projection-slice theorem \citep[][]{1986ftia.book.....B}, in our context, states that the two-dimensional Fourier transform of the Radon projection \citep[][]{radon}, along straight lines,  of 
a three-dimensional medium equals a two-dimensional slice (perpendicular to the projection direction) through the three-dimensional Fourier transform of the object. 
We now note that the phase-screen, up to a constant, is in fact a two-dimensional projection of the refractive index 
$\delta n(\vc{r})$ in three dimensions. Hence the Fourier transform of the phase-screen is simply a two-dimensional 
slice through the three-dimensional Fourier transform of the refractive index $\delta n(\vc{r})$ of the medium. The 
autocorrelation of the phase-screen as measured in the $uv$-plane is then a slice through the three dimensional
power spectrum of the refractive index, hence that of the electron density distribution. We immediately see the 
connection to the discussion in Section 5 and how this connects to Eqn.(\ref{eqn:scattering_psf}). Since point-sources in different directions project the three dimensional
power-spectrum differently on the $uv$-plane, they probe different slices through the power-spectrum. Disregarding 
the geometric curvature terms, this slice is the tangent plane to the Ewald sphere of influence at the point $\vc{s}-\vc{s}_{0}$
which for small vectorial differences are all slices through $\vc{s}-\vc{s}_{0} =(0,0,0)$, which is exactly the Fourier projection-slice theorem.

Observing over
a wide field of view allows one to build of a three-dimensional electron density power-spectrum from these different
slices. Obviously the measured auto-correlation of the phases are the result of many point sources and need to 
be disentangled. This was discussed in Section 5 in detail and is identical in the current situation. Thence, the 
phase-screen approach extended to three dimensions is completely identical to the approach taken in this paper.

\section{Results and Conclusions} 
 
A tomographic method has been introduced that allows to quantify the three-dimensional power-spectrum of the 
ionospheric electron-density fluctuations based on radio-interferometric observations by a 
two-dimensional planar array. The goal has been to provide a more complete and physically intuitive description 
of the effects of the full three dimensional ionosphere over a wide field of view on radio-interferometric images,
without any approximations about either a small field of view and/or a thin ionospheric slab (i.e.\ phase screens). 
Neither is expected to be sufficiently accurate in upcoming high dynamic range observations with low frequency arrays. 
The description is valid to first-order Born approximation, which holds
well for frequencies well above the plasma frequency of the ionosphere. Second order corrections are
typically several orders of magnitude below first order corrections at frequencies $\ga 100$~MHz.  However, we stress that the method is 
{\sl only} valid for weak scattering and that higher order corrections
are needed if the refractive index approaches unity.  The main results and conclusions are:

\begin{enumerate}

\item When modeling the ionosphere, it should not be the ionospheric electron density distribution that is 
is primary structure to model, but its autocorrelation function or equivalent its power-spectrum. Any information
about the phase structure of the ionospheric electron density distribution is lost in the cross-correlation
of the electric field when obtaining the visibilities.

\item A three-dimensional ionosphere causes a spatially
varying convolution of the sky (at second order level), whereas a two-dimensional phase-screen or a thin ionosphere results in a spatially 
invariant convolution. Ionospheric structure in the 
$w$-direction causes, to lowest order, radial stretching or squeezing of the ionospheric scattering point spread function but
leaves its tangential structure invariant. Correcting for the thickness of the ionosphere can
thus be reduced (to lowest order) to determining a single number, i.e.\ the level of stretching or squeezing.

\item Residual speckle, which can not be corrected in short time integrations, causes a diffuse
intensity halo around bright sources beyond a certain distance from the source. Whereas
longer integrations probe more of the total scattered flux of the \ipsf, because of its very steep intensity decline away from 
the source, these longer integrations also cause smearing of the instantaneous speckle pattern due to variations of 
the ionosphere. This halo (``seeing'') and related speckle noise might therefore pose a 
fundamental limitation on the ability to reach the thermal noise level in interferometers at very low frequencies after long
total integrations. 

\item Long baselines substantially help in correcting for the effects of the largest 
wave modes of the ionosphere, as seen inside the field of view of typical interferometers, as well
as for its three dimensional structure in the $w$-direction. The reason is that these large scale modes cause
a small-scale diffraction pattern and thus show up in the image on small angular scales that can only 
be resolved on long baselines (i.e.\ of a size equal or larger than the projected scale of the wave mode).  

\item An exact mathematical expression is derived that provides the power-spectrum of the ionospheric electron-density 
fluctuations from a rescaled scattered intensity field and an incident intensity field convolved with 
a complex unit phasor defined on the full sky pupil plane. This is related to a {\sl holographic principle}.
In the limit of a  small field of view, the method reduces to the usual thin-phase screen approximation. 
It is also shown, through the application of a Radon projection and the Fourier projection-slice theorem that the extension 
of the phase-screen approach to three dimensions is identical to the introduced tomographic method. 

\end{enumerate}

Whereas in this paper no direct implementation of an algorithm is given on {\sl how} to calibrate the ionosphere, a more physically intuitive
picture was presented that goes beyond the single (or multiple) phase screen models of the ionosphere and is valid in the presence of a full
three dimensional ionosphere and full-sky field of view. A mathematical expression was presented that shows
that the ionosphere causes a spatially varying point spread function (at the second order level) over the field of view, determined by the 
instantaneous three-dimensional power spectrum of the ionospheric electron density fluctuations. Subsequently
this expression was inverted, showing that through a ``holographic'' principle one can extract information about
the three-dimensional structure of the ionosphere from only electric field measurements on a two dimensional plane.
In forthcoming work it remains to be investigated whether these results can be implemented in
a practical algorithm in the context of the measurement equation 
\citep[see e.g.][]{2009arXiv0911.3942M} 
in order to calibrate visibilities {\sl without} assumptions about thin ionospheric phase screens, and by using instead 
the full three dimensional power-spectrum of the ionosphere (i.e.\ ``tomographic self-calibration'').

\acknowledgements

L.K. is supported through an NWO-VIDI program subsidy. The author also acknowledges
Saleem Zaroubi, Rajat Thomas and Ger de Bruyn for very useful and clarifying discussions.

\bibliographystyle{apj}

\begin{thebibliography}{31}
\expandafter\ifx\csname natexlab\endcsname\relax\def\natexlab#1{#1}\fi

\bibitem[{{Born} \& {Wolf}(1999)}]{1999prop.book.....B}
{Born}, M., \& {Wolf}, E. 1999, {Principles of Optics}, ed. {Born, M.~\& Wolf,
  E.}

\bibitem[{{Bourgois}(1981)}]{1981A&A...102..212B}
{Bourgois}, G. 1981, \aap, 102, 212

\bibitem[{{Bracewell}(1986)}]{1986ftia.book.....B}
{Bracewell}, R.~N. 1986, {The Fourier Transform and its applications}, ed.
  {Bracewell, R.~N.}

\bibitem[{{Bramley}(1954)}]{1954RSPSA.225..515B}
{Bramley}, E.~N. 1954, Royal Society of London Proceedings Series A, 225, 515

\bibitem[{{Carozzi} \& {Woan}(2009)}]{2009MNRAS.395.1558C}
{Carozzi}, T.~D., \& {Woan}, G. 2009, \mnras, 395, 1558

\bibitem[{{Cohen} \& {R{\"o}ttgering}(2009)}]{2009AJ....138..439C}
{Cohen}, A.~S., \& {R{\"o}ttgering}, H.~J.~A. 2009, \aj, 138, 439

\bibitem[{{Cornwell} \& {Perley}(1992)}]{1992A&A...261..353C}
{Cornwell}, T.~J., \& {Perley}, R.~A. 1992, \aap, 261, 353

\bibitem[{{Cronyn}(1972)}]{1972ApJ...174..181C}
{Cronyn}, W.~M. 1972, \apj, 174, 181

\bibitem[{{Deans}(1983)}]{radon}
{Deans}, S.~R. 1983, New York: John Wiley \& Sons

\bibitem[{{Ewald}(1969)}]{1969AcCrA..25..103E}
{Ewald}, P.~P. 1969, Acta Crystallographica Section A, 25, 103

\bibitem[{{Goodman}(1985)}]{1985stop.book.....G}
{Goodman}, J.~W. 1985, {Statistical Optics}, ed. {Goodman, J.~W.}

\bibitem[{{Hamaker} {et~al.}(1996){Hamaker}, {Bregman}, \&
  {Sault}}]{1996A&AS..117..137H}
{Hamaker}, J.~P., {Bregman}, J.~D., \& {Sault}, R.~J. 1996, \aaps, 117, 137

\bibitem[{{Hewish}(1951)}]{1951RSPSA.209...81H}
{Hewish}, A. 1951, Royal Society of London Proceedings Series A, 209, 81

\bibitem[{{Hewish}(1952)}]{1952RSPSA.214..494H}
---. 1952, Royal Society of London Proceedings Series A, 214, 494

\bibitem[{{Intema} {et~al.}(2009){Intema}, {van der Tol}, {Cotton}, {Cohen},
  {van Bemmel}, \& {R{\"o}ttgering}}]{2009A&A...501.1185I}
{Intema}, H.~T., {van der Tol}, S., {Cotton}, W.~D., {Cohen}, A.~S., {van
  Bemmel}, I.~M., \& {R{\"o}ttgering}, H.~J.~A. 2009, \aap, 501, 1185

\bibitem[{{Jacobson} \& {Erickson}(1992)}]{1992A&A...257..401J}
{Jacobson}, A.~R., \& {Erickson}, W.~C. 1992, \aap, 257, 401

\bibitem[{{Kravtsov}(1992)}]{1992RPPh...55...39K}
{Kravtsov}, Y.~A. 1992, Reports on Progress in Physics, 55, 39

\bibitem[{{Liu} {et~al.}(2010){Liu}, {Tegmark}, {Morrison}, {Lutomirski}, \&
  {Zaldarriaga}}]{2010arXiv1001.5268L}
{Liu}, A., {Tegmark}, M., {Morrison}, S., {Lutomirski}, A., \& {Zaldarriaga},
  M. 2010, ArXiv e-prints

\bibitem[{{Lonsdale}(2005)}]{2005ASPC..345..399L}
{Lonsdale}, C.~J. 2005, in Astronomical Society of the Pacific Conference
  Series, Vol. 345, Astronomical Society of the Pacific Conference Series, ed.
  {N.~Kassim, M.~Perez, W.~Junor, \& P.~Henning}, 399--+

\bibitem[{{Lutomirski} \& {Yura}(1971)}]{1971ApOpt..10.1652L}
{Lutomirski}, R.~F., \& {Yura}, H.~T. 1971, \ao, 10, 1652

\bibitem[{{Mandel} \& {Wolf}(1965)}]{1965RvMP...37..231M}
{Mandel}, L., \& {Wolf}, E. 1965, Reviews of Modern Physics, 37, 231

\bibitem[{{Matejek} \& {Morales}(2009)}]{2009arXiv0911.3942M}
{Matejek}, M.~S., \& {Morales}, M.~F. 2009, ArXiv e-prints

\bibitem[{{Narayan}(1992)}]{1992RSPTA.341..151N}
{Narayan}, R. 1992, Royal Society of London Philosophical Transactions Series
  A, 341, 151

\bibitem[{{Pearson} \& {Readhead}(1984)}]{1984ARA&A..22...97P}
{Pearson}, T.~J., \& {Readhead}, A.~C.~S. 1984, \araa, 22, 97

\bibitem[{{Racine} {et~al.}(1999){Racine}, {Walker}, {Nadeau}, {Doyon}, \&
  {Marois}}]{1999PASP..111..587R}
{Racine}, R., {Walker}, G.~A.~H., {Nadeau}, D., {Doyon}, R., \& {Marois}, C.
  1999, \pasp, 111, 587

\bibitem[{{Ratcliffe}(1956)}]{1956RPPh...19..188R}
{Ratcliffe}, J.~A. 1956, Reports on Progress in Physics, 19, 188

\bibitem[{{Salpeter}(1967)}]{1967ApJ...147..433S}
{Salpeter}, E.~E. 1967, \apj, 147, 433

\bibitem[{{Sramek} \& {Schwab}(1989)}]{1989ASPC....6..117S}
{Sramek}, R.~A., \& {Schwab}, F.~R. 1989, in Astronomical Society of the
  Pacific Conference Series, Vol.~6, Synthesis Imaging in Radio Astronomy, ed.
  {R.~A.~Perley, F.~R.~Schwab, \& A.~H.~Bridle}, 117--+

\bibitem[{{Thompson} {et~al.}(2001){Thompson}, {Moran}, \&
  {Swenson}}]{2001isra.book.....T}
{Thompson}, A.~R., {Moran}, J.~M., \& {Swenson}, Jr., G.~W. 2001,
  {Interferometry and Synthesis in Radio Astronomy, 2nd Edition}, ed.
  {Thompson, A.~R., Moran, J.~M., \& Swenson, G.~W., Jr.}

\bibitem[{{Weyl}(1919)}]{1919AnP...365..481W}
{Weyl}, H. 1919, Annalen der Physik, 365, 481

\bibitem[{{Wolf}(1969)}]{1969OptCo...1..153W}
{Wolf}, E. 1969, Optics Communications, 1, 153

\end{thebibliography}

\end{document}